\documentclass{osa-article}

\journal{osajournal}

\usepackage{amsmath,bm}
\usepackage{mathtools}
\usepackage{booktabs}
\usepackage{multirow}
\articletype{Research Article}

\begin{document}

\title{Compressive Spectral Image Reconstruction using Deep Prior and Low-Rank Tensor Representation}

\author{Jorge Bacca,\authormark{} Yesid Fonseca,\authormark{} and Henry Arguello\authormark{*}}

\address{\authormark{}Department of Systems Engineering, Universidad Industrial de Santander, Bucaramanga, Colombia}

\email{\authormark{*}henarfu@uis.edu.co}


\begin{abstract}
Compressive spectral imaging (CSI) has emerged as an alternative spectral image acquisition technology, which reduces the number of measurements at the cost of requiring a recovery process.  In general, the reconstruction methods are based on hand-crafted priors used as regularizers in optimization algorithms or recent deep neural networks employed as an image generator to learn a non-linear mapping from the low-dimensional compressed measurements to the image space. However, these \textcolor{black}{deep learning} methods need many spectral images to obtain good performance. In this work, a deep recovery framework for CSI without training data is presented. The proposed method is based on the fact that the structure of some deep neural networks and an appropriated low-dimensional structure are sufficient to impose a structure of the underlying spectral image from CSI. We analyzed the low-dimension structure via the Tucker representation, modeled in the first net layer. The proposed scheme is obtained by minimizing the $\ell_2$-norm distance between the compressive measurements and the predicted measurements, and the desired recovered spectral image is formed just before the forward operator. Simulated and experimental results verify the effectiveness of the proposed method \textcolor{black}{for the coded aperture snapshot spectral imaging.}
\end{abstract}

\section{Introduction}
Spectral imaging (SI) deals with capturing the spatial information of a target in a broader range of the electromagnetic spectrum compared to a conventional RGB imaging system. This additional information is useful for some applications such as biomedical imaging \cite{lu2014medical}, crop identification \cite{zhang2012tensor}, and surveillance~\cite{yuen2010introduction}. SI can be denoted as a 3D tensor $\bm{\mathcal{X}}\in \mathbb{R}^{M \times N \times L}$ with $M\times N$ as the spatial pixels and $L$ spectral bands \cite{zhang2012tensor}. Traditional methods to acquire SI are based on scanning along one of its tensor modes, which results in time-consuming systems, and therefore, prohibits its usage in dynamic scenes \cite{hinojosa2018coded}.

Alternatively, based on the compressive sensing (CS) theory, new imaging snapshots systems acquire 2D multiplexed projections of a scene instead of directly acquire all voxels, resulting in an image compression via hardware \cite{arce2014compressive}. To date, different compressive spectral imaging (CSI) techniques have been proposed \cite{cao2016computational,correa2016multiple,arguello2014colored,wagadarikar2008single,gehm2007single,shauli2020dual,baek2017compact,wang2018compressive,kar2019compressive,monakhova2020spectral}. \textcolor{black}{This work focuses on the pioneer} coded aperture snapshot spectral imaging (CASSI) system \cite{gehm2007single}, which uses optical elements to encode and disperse the incoming light to acquire 2D intensity projections. Even though CSI yield efficient sensing, a reconstruction process from the compressed measurements is needed, since it  results in finding a solution to an under-determined system~\cite{arce2014compressive}. This recovery problem is addressed by representing the 3D scene as a 1D vector and assuming particular spectral image nature priors in different dimensions used as regularization in an optimization problem  \cite{hinojosa2018coded,zhang2019computational}. For instance, \cite{kittle2010multiframe,wang2015dual} assume low total variation, \cite{wagadarikar2008single,correa2016multiple} explore the sparsity assumption of the scene in some orthogonal basis, \cite{fu2016exploiting,wang2016adaptive} use non-local similarity, and \cite{bacca2019noniterative,gelvez2017joint} employ low-rank structures. However, these hand-crafted priors do not often represent the wide variety and non-linearity of spectral images, and the vectorization ignores the high-dimensional structure of the scene, resulting in low reconstruction quality~\cite{wang2019hyperspectral}. 

On the other hand,  \textcolor{black}{deep learning} recovery methods are based on the power of the deep neural networks as image generators, where the goal is to learn a non-linear transformation that maps a low-dimensional feature into realistic spectral images~\cite{hyder2019generative}. In particular, with a vast spectral data set,~\cite{wang2018hyperreconnet,xiong2017hscnn,miao2019lambda,gedalin2019deepcubenet} learn inverse networks that map the low-dimensional compressed measurements to the desired spectral image~\cite{bacca2020coupled}. These methods have shown high performance speed and reconstrucion quality. However, they are very dependent on training data, and small variations in the sensing system would require re-training of the model \cite{wang2019hyperspectral}.   Alternative solutions such as~\cite{choi2017high},  take the sensing model into account  when solving an optimization problem where the prior is learned using convolutional auto-encoder with a spectral data set,  \cite{choi2017high,wang2019hyperspectral,zhang2019hyperspectral,wang2020dnu,sogabe2020admm} use unrolled-based methods, \textcolor{black}{which are networks inspired by optimization algorithms,} where the prior is intrinsically learned, \textcolor{black}{or more recently, auhors in \cite{bacca2020coupled,meng2020end,oktem2020high} learn the sensing matrix jointly through end-to-end optimization.} Although these methods have proven to be more general, they still depend on training data.

In this paper, a deep recovery framework for reconstructing spectral images from CSI measurements without training data requirements is proposed.  The method is based on the fact that the deep convolutional neural networks and the appropriated low-dimensional input are sufficient to learn/generate the image representation without any training data, and therefore, to recover a spectral image directly from the CSI measurements. In particular, the proposed method designs a deep neural network, \textcolor{black}{where the network input is also learned by imposing a low-dimensional 3D tensor commonly used in SI, which is then refined by convolutional operations to generate the non-linearity recovered SI.}  The weights of this neural network are randomly initialized and fitted to guarantee that the reconstruction suits the CSI measurement via $\ell_2$-norm minimization over the CSI measurement; therefore, the recovered image is formed just before the forward operator. The proposed method is expressed as an end-to-end optimization by modeling the forward compressive sensing model as a non-trainable layer; consequently, it can be solved using any deep learning algorithm like stochastic gradient descent. Additionally, we analyzed the importance of the low-dimensional tensor structure in the first layer via low-rank Tucker representation, which imposes a low-rank 3D-prior. Since there is no more information available other than the compressive spectral measurements, the proposed method is more related to hand-crafted techniques. Results in simulated and real data \textcolor{black}{of a  CASSI system as CSI}  demonstrate that the proposed method outperforms the hand-crafted methods in many scenarios and obtains comparable results with \textcolor{black}{deep learning}~approaches.

\section{Related work}

\subsection{Hand-Crafted CS Reconstruction}The traditional CS recovery algorithms are considered hand-designed since they use some expert knowledge of the signal, known as a signal prior  \cite{choi2017high}. These methods are based on optimization techniques that design a data fidelity term, and incorporate the prior as a regularization term \cite{figueiredo2007gradient}. The most common prior is assuming that the signal is sparse on a given basis, such as Wavelet \cite{candes2008introduction}, discrete cosine transform (DCT)\cite{arce2014compressive}, among others\cite{arce2014compressive}. This sparsity assumption is imposed in different methods by applying $\ell_0$ or $\ell_1$ regularizers. Examples of algorithms that use sparsity priors include, the GPSR  \cite{figueiredo2007gradient}, ADMM \cite{boyd2011distributed}, CSALSA\cite{csalsa}, ISTA \cite{daubechies2004iterative}, AMP \cite{donoho2009message} among others. In CSI, some specific kinds of prior are used. For instance, \cite{wagadarikar2008single} assumes low total variation, \cite{correa2016multiple} explores the spatial sparsity assumption of the scene in Wavelet domain, and the spectral sparsity in the DCT domain \cite{fu2016exploiting,wang2016adaptive}; furthermore, \cite{bacca2019noniterative,gelvez2017joint} employ low-rank structures based on the linear mixture model. Exploring tensor structure, low-rank tensor recovery methods have been also proposed \cite{zhang2019computational,yang2015compressive}.  However, these hand-crafted methods require expert knowledge of the target to select which prior to use. Therefore, they do not represent the wide variety and the non-linearity of spectral image representations.

\subsection{\textcolor{black}{CS Recovery methods based on Deep Learning}}
\textcolor{black}{Deep learning (DL) methods for CS} are based on learning a non-linear inverse mapping from the compressive measurements to a realistic image. In particular, with a vast dataset of ground-truth and compressive measurement pairs, these methods are used to learn a non-linear network by minimizing the distance between the output of the net and the ground-truth.  The main difference between the state-of-the-art methods is their network architecture. For instance, \cite{mousavi2015deep} learns a stacked auto-encoder, convolution layers are applied in \cite{mousavi2017learning}, and convolutional, residual, and fully-connected layers are also used in ~\cite{dave2018solving,palangi2016distributed,yao2019dr2,kulkarni2016reconnet}. In particular, for CSI, \cite{xiong2017hscnn} was the first work that used a \textcolor{black}{deep learning} approach, where, an initialization obtained from TwiST \cite{bioucas2007new} was refined using denoising networks; \cite{wang2019hyperspectral} proposed a particular model to explore the spatial and spectral information and to design the coded aperture usually included in CSI architectures. Furthermore, based on the structure of the U-net, \cite{gedalin2019deepcubenet} proposed a non-linear mapping replacing the 2D for 3D convolutions, and \cite{miao2019lambda} developed a generative model based on the U-net. These methods have shown high performance in reconstruction quality, and once trained, they allow real-time reconstruction. However, these approaches are highly dependent on the data-set used. Furthermore, small-variations in the compressive measurements, such as type of noise or changes in the sensing matrix, would require a time-consuming re-training.

Recently, some works have considered the sensing model to proposed a mixed approach which considers the hand crafted as well as the \textcolor{black}{deep learning} CS reconstruction. In particular, these methods use a deep network or denoiser to replace the hand-crafted prior, then, this non-linear prior is employed in the optimization algorithm~\cite{dave2018solving}. For instance, Plug-and-play priors (PnP) use pre-existing denoisers as a proximal step~\cite{yuan2020plug,chan2016plug},  \cite{rick2017one} learns the proximal mapping using a convolutional network, and~\cite{choi2017high} learns a SI prior, through a convolutional autoencoder, which is then incorporated into the optimization problem. More recently, D-AMP~\cite{metzler2017learned}, ISTA-Net~\cite{zhang2018ista}, ADMM-Net~\cite{sun2016deep}, and DNU~\cite{wang2020dnu} use the unrolled based method that incorporates the optimization steps into the deep network architecture using residual networks; consequently, they can learn the prior and the parameters via end-to-end training. This strategy is also employed for CSI in~\cite{wang2019hyperspectral,zhang2019hyperspectral}. Although these methods have proven to be more general, they still depend on training data, which is limited in SI.	

\subsection{Deep Image Prior using Generative Model}

The generative model (GM) has been used for CS recovery~\cite{bora2017compressed}. The goal in GM is to generate a realistic image from a low-dimensional
latent input. For instance, \cite{bora2017compressed,wu2019deep} use a pre-trained deep neural network and obtain the low-dimensional input, which minimizes the distance between the compressive measurements and the output of the net. On the other hand, \cite{ulyanov2018deep} shows that a pre-trained network is not necessary. Instead of finding the low-dimensional latent input, \cite{ulyanov2018deep} uses a fixed random variable as latent input, then the weights of the model are updated to obtain an optimal result. The drawback of this method is its sensitivity to changes in the application, the fixed input or the network architecture, which usually require small random disturbances to obtain a good performance.\textcolor{black}{ The proposed method in this work is closely related to  \cite{ulyanov2018deep,wu2019deep}, where the parameters of the network model are optimized, but instead of remaining fixed the network input, we also optimized it in an end-to-end approach imposing a low-dimensional representation (based on a Tucker representation, which is helpful for SI) for a CSI architecture, which restricts the feasible set, showing better performance as is presented in the simulation section.}

\subsection*{Notation:} Through the paper, vectors are represented with boldface lowercase letters, e.g., $\bm{x},$ and matrices are denoted as boldface capital letters $\mathbf{X}$. The 3D tensors are denoted as $\bm{\mathcal{X}}\in\mathbb{R}^{M\times N\times L}$ and	the $1$-mode product of a tensor $\bm{\mathcal{X}}_o\in\mathbb{R}^{M_p\times N_p\times L_p}$ with a matrix $\mathbf{U}\in\mathbb{R}^{M\times M_p}$ is written as $\bm{\mathcal{X}}=\bm{\mathcal{X}}_o\times_1 \mathbf{U}$ where $\bm{\mathcal{X}}\in\mathbb{R}^{M\times N_p\times L_p}$, and \vspace{-0.5em}
\begin{align*}
\bm{\mathcal{X}}_{(m,n,\ell)}=\sum_{\hat{m}=1}^{M_p}\mathbf{U}_{(m,\hat{m})}\bm{\mathcal{X}}_{o(\hat{m},n,\ell)}.
\end{align*}
In the same way, the 2-mode and 3-mode products can be defined. We introduce the function
\label{Sec:Notation}
$\text{shift}_{\ell}(\cdot):\mathbb{R}^{M\times N}\rightarrow \mathbb{R}^{M\times (N+L-1)}$ which refers to a shifting operator, i.e., for a given $\mathbf{X}$ we have that
\begin{align*}
\text{shift}_{\ell}(\mathbf{X}):=\begin{cases}
\mathbf{X}_{(m,n-\ell)}, &\text{ if } 1\leq n-l \leq N\\
0, & \text{ otherwise}.
\end{cases}
\end{align*}
Finally, the function $\text{vect}(\cdot):\mathbb{R}^{M\times N\times L} \rightarrow \mathbb{R}^{MNL}$ represents the vectorization of a tensor.

	\section{Compressed Measurements Acquisition}
\begin{figure}[!b]
	\centering
	\includegraphics[width=0.8\linewidth]{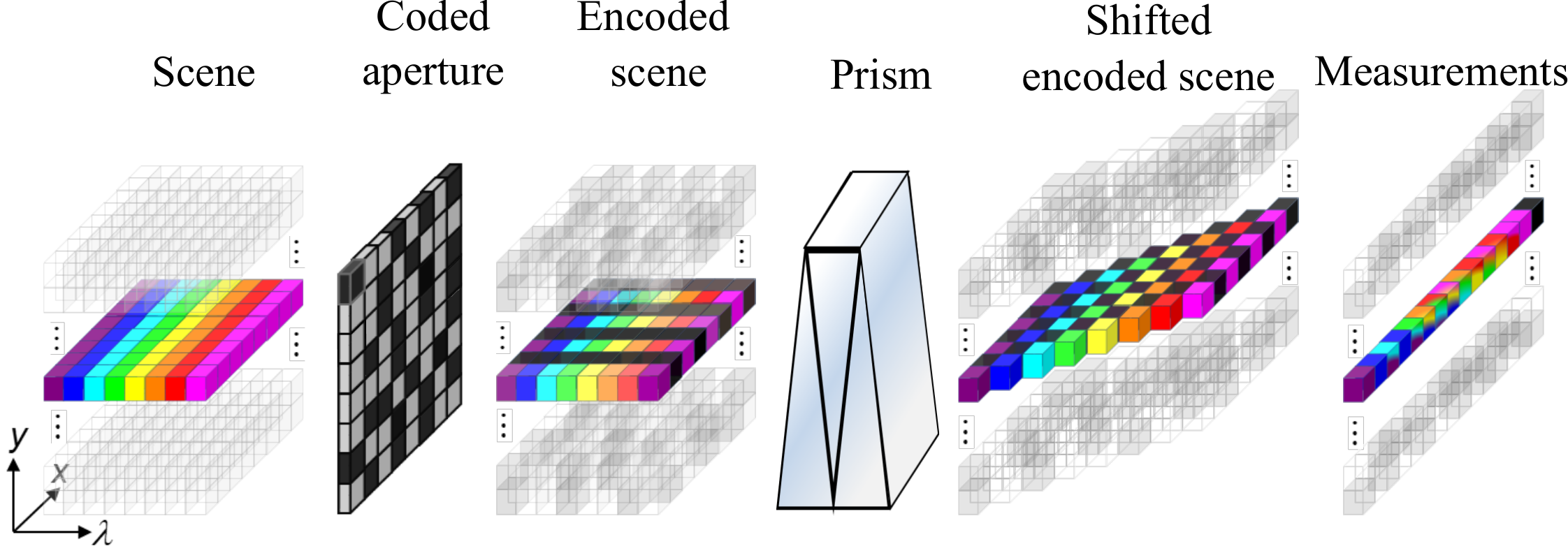}\vspace{-0.5em}
	\caption{Physical sensing phenomena in CASSI, which is the CSI prototype used to validate the proposed approach.}
	\label{fig:3DCASSI_sensingApp}
\end{figure}
The CASSI sensing approach is used in order to acquire the compressed measurements of a spectral scene \cite{gehm2007single}. This architecture is composed of three main optical elements: a coded aperture, a prism as a dispersive element, and a gray-scale detector, as illustrated in Fig \ref{fig:3DCASSI_sensingApp}. The spatial-spectral data cube is represented as $\bm{\mathcal{X}}\in\mathbb{R}^{M\times N\times L}$ with $M\times N$ spatial dimensions, $L$ spectral bands, and $\mathbf{X}_{\ell}\in\mathbb{R}^{M\times N}$ denotes the 2D spectral intensity image of $\bm{\mathcal{X}}$ at the $\ell$-th spectral band. As shown in Fig. \ref{fig:3DCASSI_sensingApp}, each spatial position of the scene is modulated by a coded aperture $\mathbf{C}\in\{0,1\}^{M\times N}$, which block/unblock the incoming light, then, the coded spectral scene passes through the prism creating a horizontal shifting. Finally, the coded shifted spectral scene is integrated along the spectral axis by the detector, resulting in the 2D compressed measurement $\mathbf{Y}\in\mathbb{R}^{M\times (N+L-1)}$. In CSI, it is possible to acquire $S<L$ different measurement snapshots of the same spectral data cube employing $S$ different patterns in the coded aperture. Therefore, the output of the sensing process at the $s$-th spectral snapshot can be mathematically expressed as
\begin{equation}
\mathbf{Y}^{(s)}=\sum_{\ell=1}^{L}\text{shift}_{\ell-1}\left( \mathbf{X}_{\ell}\odot\mathbf{C}^{(s)}\right),
\label{eq:discretizedOutput}
\end{equation}
where the $\ell$-th spectral band, $\mathbf{X}_\ell$, of the tensor $\bm{\mathcal{X}}$ is shifted with the operator $\text{shift}_{\ell-1}(\cdot)$, and $\odot$ denotes the element-wise product with the 2D coded aperture $\mathbf{C}^{(s)}$.

The CASSI sensing model can be seen as a linear operator, after stacking the measurements of multiple shots as $\bm{y}=[\text{vect}(\mathbf{Y}^{(1)})^T, \cdots \text{vect}(\mathbf{Y}^{(S)})^T]$. Thus, the system matrix model can be expressed as
\begin{equation}
\bm{y}=\mathbf{H}\text{vect}(\bm{\mathcal{X}}),
\label{eq:measuresForClustering}
\end{equation}
where $\mathbf{H}\in \mathbb{R}^{SM(N+L-1) \times MNL}$ represents the linear sensing matrix of CASSI.

\section{Compressive Spectral Reconstruction} \label{sec:Design}

\begin{figure}[!t]
	\centering
	\includegraphics[width=1\linewidth]{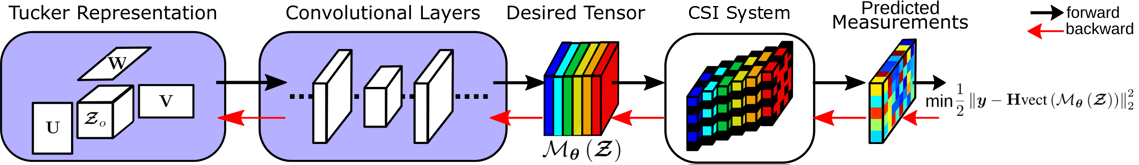}\vspace{0.3em}
	\caption{Visual representation of the proposed deep neural scheme, where the boxes with background color represent the learning parameters, the white box stand for the non-trainable CSI system, and the non-box blocks represent the outputs of the layers. }
	\label{fig:proposed_method}
\end{figure}

The goal in CSI is to recover the spectral image $\boldsymbol{\mathcal{X}}\in \mathbb{R}^{M \times N \times L}$  from the compressive measurements $\bm{y}$. Since $SM(N+L-1)\ll MNL$, this problem consists in solving an undetermined system, which is addressed by restricting the feasible set of solutions using image priors as regularizers. A tensor formulation for addressing this problem is described below
\begin{align}
\label{eq:traditional_reconstruction}
\underset{ \bm{\mathcal{Z}}_o \in\mathbb{R}^{M\times N\times L}  }{\mbox{minimize}} \hspace{3mm} &  \frac{1}{2}\left\| \bm{y}-\mathbf{H}\text{vect}\left( \bm{\mathcal{X}}\right) \right\|^2_2+\lambda\cdot\phi(\bm{\mathcal{Z}'}_o) \\ \nonumber \text{subject to}  \hspace{7mm}  &  \bm{\mathcal{X}}=\bm{\mathcal{Z}'}_o\times_1\mathbf{U}'\times_2\mathbf{V}'\times_3\mathbf{W}',
\end{align}
where the matrices $\mathbf{U}'\in\mathbb{R}^{M\times M},\mathbf{V}'\in\mathbb{R}^{N\times N}$ and $\mathbf{W}'\in\mathbb{R}^{L\times L}$ are fixed and known orthogonal matrices, which usually are the matrix representation of the Wavelet and the Discrete Cosine transforms; $\bm{\mathcal{Z}'}_o$ is the representation of the spectral image in the given basis and $\phi(\cdot): \mathbb{R}^{M\times N\times L} \rightarrow \mathbb{R} $ is a regularization function that imposes particular image priors with $\lambda$ as the regularization parameter~\cite{figueiredo2007gradient}.

Unlike the hand-craft priors as sparsity \cite{arce2014compressive}, we explore the power of some deep neural networks as image generators that map a low-dimensional feature tensor $\bm{\mathcal{Z}}\in\mathbb{R}^{M\times N\times L}$ to the image as
\begin{equation}
\bm{\mathcal{X}} = \mathcal{M}_{\bm\theta}(\bm{\mathcal{Z}}),
\end{equation}
where $\mathcal{M}_{\bm\theta}(\cdot)$ represents a deep network, with ${\bm\theta}$ as the net-parameters. To ensure a low-dimensional structure over the feature tensor, this work used the Tucker representation, i.e.,  $\bm{\mathcal{Z}}=\bm{\mathcal{Z}}_o\times_1\mathbf{U}\times_2\mathbf{V}\times_3\mathbf{W}$ with  $\bm{\mathcal{Z}}_o\in\mathbb{R}^{M_{\rho}\times N_{\rho}\times L_{\rho}}$ as a 3D low dimensional tensor, with $M_\rho<M$, $N_\rho<N$ and~$L_\rho<L$.
This representation, in the input of the network, aims to maintain the 3D structure of the spectral images, exploits the inherent low-rank of this data~\cite{wang2017compressive,leon2020online}, and also implicitly constraint the output $\boldsymbol{\mathcal{X}}$ in a low-dimensional manifold via the architecture and the weights of the net~\cite{wu2019deep}. \textcolor{black}{It is worth highlighting that, unlike~\cite{wang2017compressive,leon2020online}, we do not satisfy low-rank structure in the recovered spectral image (output of the network). Instead, we impose Tucker decomposition on the input network, which expects that after some convolution layer, extract some non-linearity features present in the SI.}

In this paper, we are focused in a blind representation, where instead of have a pre-training network or huge amount of data to train this deep neural representation, we express an optimization problem which learns the weight $\bm\theta$ in the generative network $\mathcal{M}_{\bm\theta}$ and also the tensor feature  $\bm{\mathcal{Z}}$ with its Tucker representation elements as $\bm{\mathcal{Z}}_o,\mathbf{U},\mathbf{V}$ and $\mathbf{W}$. All the parameters of this optimization problem are randomly initialized and the only  available information are the compressive measurements and the sensing model, i.e, the optimization problem is data training independent. In particular,  we explore the prior implicitly captured by the choice of the generator network structure, which is usually composed of convolutional operations, and the importance of the low-rank representation feature, therefore, the proposed method consists of solving the following optimization problem
\begin{align}
\label{eq:mainproblem}
\underset{ {\bm\theta}, \bm{\mathcal{Z}}_o, \mathbf{U,V,W}   }{\mbox{minimize}} \hspace{3mm} &  \frac{1}{2}\left\| \bm{y}-\mathbf{H}\text{vect}\left( \mathcal{M}_{{\bm\theta}}\left(\bm{\mathcal{Z}}\right)\right) \right\|^2_2  \\ \nonumber \text{subject to}  \hspace{7mm}  &  \bm{\mathcal{Z}}=\bm{\mathcal{Z}}_o\times_1\mathbf{U}\times_2\mathbf{V}\times_3\mathbf{W},
\end{align}
where the recovery is  $\bm{\mathcal{X}}^{*}= \mathcal{M}_{\bm{\theta}^{*}}(\bm{\mathcal{Z}}_o^*\times_1\mathbf{U}^*\times_2\mathbf{V}^*\times_3\mathbf{W}^*)$. This optimization problem can be solved using an end-to-end neural network framework, as shown in Fig. \ref{fig:proposed_method}. In this way, the input, that is common in all neural networks, is replaced with a custom layer with $\bm{\mathcal{Z}}_o,\mathbf{U, V, W}$ as learnable parameters, which construct the low-rank Tucker representation of $\bm{\mathcal{Z}}$, then this tensor $\bm{\mathcal{Z}}$ is refined with convolutional layers via $\mathcal{M}_{\bm \theta}(\bm{\mathcal{Z}})$; these optimization variables are represented by the first two blue-blocks in the Fig. \ref{fig:proposed_method}. The final layer in the proposed method is a non-training layer which models the forward sensing operator $\mathbf{H}\text{vect}\left( \mathcal{M}_{{\bm\theta}}\left(\bm{\mathcal{Z}}\right)\right)$ to obtain the compressive measurements $\bm{y}$ as the output of the net.  Therefore, the problem in \eqref{eq:mainproblem} can be solved with  state-of-the-art deep learning optimization algorithm, such as, stochastic gradient descent. Once the parameters are optimized, the desired SI is recovered just before the non-trainable layer labeled as "CSI system" in Fig. \ref{fig:proposed_method}.

\section{Simulation and Results}
\begin{figure}[!b]
	\centering
	\includegraphics[width=1\linewidth]{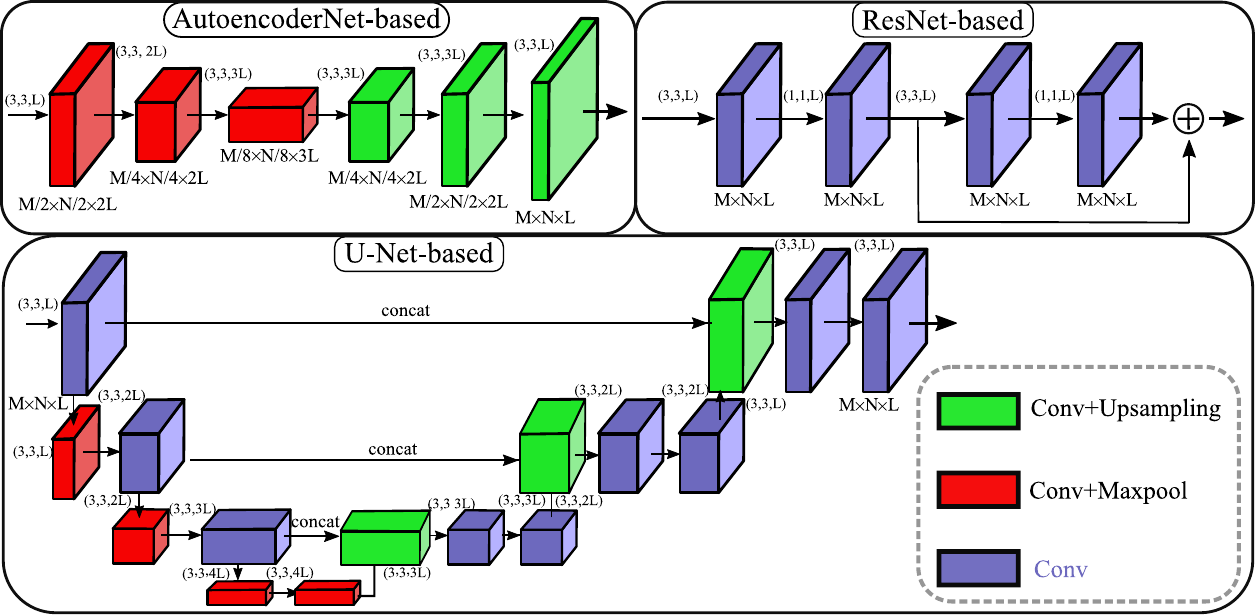} 
	\caption{Visual representation of the three network models used: U-Net-based, AutoencoderNet-based and ResNet-based. The color represents the different layers in each network. }
	\label{fig:Network_architeuctre}
\end{figure}
In this section, the performance of the proposed compressive spectral image reconstruction approach is presented. The performance metrics used are the peak-signal-to-noise ratio (PSNR)~\cite{arce2014compressive}, the structural similarity (SSIM)~\cite{wang2004image}, and \textcolor{black}{the spectral angle mapping (SAM)~\cite{bacca2019noniterative} between two spectral signature calculated as \begin{equation}
    \text{SAM} = \cos^{-1}\left(\frac{\mathbf{f}_1^T\mathbf{f}_2}{||\mathbf{f}_1||_2.||\mathbf{f}_2||_2}\right).
\end{equation}} PSNR and SSIM are calculated as the average of each 2D spatial image through the bands, and the SAM is the average of all spectral pixels. Four different tests are presented to validate the proposed method. The first test evaluates the importance of the low-rank tensor representation; the second test compares the recovery of the \textcolor{black}{deep learning} methods with the proposed approach; the third evaluates the proposed method in different noisy scenarios and for a different number of shots against the non-data dependent state-of-the-art algorithms, \textcolor{black}{where the measurements were corrupted at different level of signal-to-noise (SNR) ratios calculated as
\begin{equation}
    \text{SNR}  = 20 \log_{10} \left(\frac{||\mathbf{y}||_{2}}{||\mathbf{w}||_{2}}\right),
\end{equation} where $\mathbf{w}$ stands for the additive noise.}
Finally, the proposed method is evaluated using two compressive spectral images obtained with a real test-bed implementation \footnote{The code can be find \url{https://github.com/jorgebaccauis/Deep_Prior_Low_Rank}}. \textcolor{black}{All the simulated experiments use the CASSI as CSI system with 50\% of transmittance.}


\subsection{Rank level }

\begin{figure}[!t]
	\centering
	\includegraphics[width=1\linewidth]{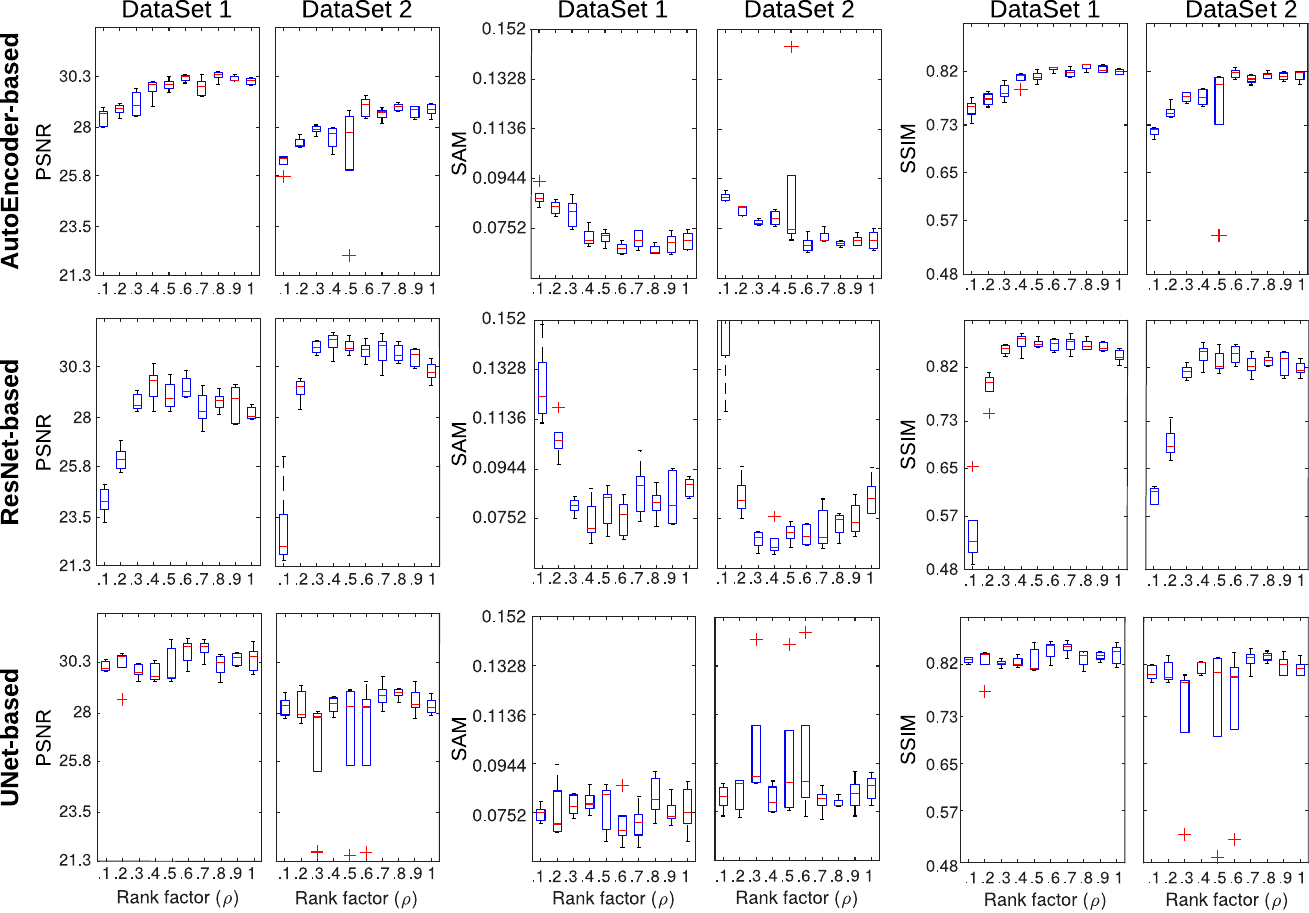}
	\caption{\textcolor{black}{PSNR, SAM and SSIM Box plots} for the different network architectures varying the rank factor $\rho$, with $5$ run trials.}
	\label{fig:Box_plot_rank}
\end{figure}

This section evaluates the importance of the rank level in the 3D tensor using the Tucker representation, which is placed at the first block of our model, as illustrated in Fig. \ref{fig:proposed_method}. For that, two spectral images with $M = 256\times N = 256$ pixels, and $L = 10$ spectral bands between $400$ and $700$nm from~\cite{marquez2020compressive} where chosen.
Three different network architectures were tested as ''Convolutional Layers'' for the second block in the Figure~\ref{fig:proposed_method}. The first network architecture is a simple ResNet-based model~\cite{he2016deep}, with a single skip connection and four convolutional layers, as shown in the Figure~\ref{fig:Network_architeuctre} with $2150$ parameters. The second architecture, also shown in Fig.~\ref{fig:Network_architeuctre}, is a convolution Autoencoder-based~\cite{masci2011stacked}, with $8160$ training parameters, and six convolutional layers. The third architecture tested and depicted in FIg.~\ref{fig:Network_architeuctre}, is a Unet-based~\cite{ronneberger2015u}, without drop-out layers, and, in the contracting part, the feature information is increased using multiples of $L=10$, i.e., $L,2L$ and $3L$ as is illustrated in Fig.\ref{fig:Network_architeuctre}, resulting in $92190$ training parameters. This test is focused on a single snapshot for a randomly coded aperture generated from a Bernoulli distribution with mean  0.5 in a noiseless scenario, \textcolor{black}{i.e., $\infty$ of SNR.}

As mentioned, the tensor feature $\bm{\mathcal{Z}}\in\mathbb{R}^{M\times N\times L}$ comes from a low-dimensional kernel $\bm{\mathcal{Z}}_o\in\mathbb{R}^{M_\rho\times N_\rho\times L_\rho}$; then, to evaluate the importance of the rank-level in the Tucker representation, we establish the following relationship
\begin{equation}
\frac{M_\rho}{M}=\frac{N_\rho}{N}=\frac{L_\rho}{L}=\rho,
\label{eq_rank_factor}
\end{equation}
where $\rho \in (0,1]$, is referred as the hyper-parameter \textit{rank factor}. Furthermore, as the parameters of the problem in  \eqref{eq:mainproblem} are randomly initialized, we simulated five realizations. The average results for this 5 realizations are summarized in the Figure \ref{fig:Box_plot_rank}. Notice that for the three network architectures and the two datasets, the rank factor is a crucial hyper-parameter to obtain a good reconstruction.\textcolor{black}{ In particular, the optimal value is $\rho=\{0.6, 0.4\}$ for the AutoeconderNet-based, and ResNet-based, and for both Datasets. The best $\rho$ parameter for the Unet-based, vary between 0.2 and 0.4 as is shown in all metrics in Fig. \ref{fig:Box_plot_rank}}. Furthermore, notice that a small value of $\rho$ presents the worst case for all the networks. Also, notice all the network configurations obtain around 30 dB, which is the best-obtained results,  for different $\rho$ values; however, the AutoencoderNet-based is more stable compared with the other networks. This result shows the importance of the low-rank tensor representation in the first layer, where the optimal value changes for each dataset and each network architecture.

\subsection{\textcolor{black}{Deep Learning} Methods Comparison}

\begin{figure}[!t]
	\centering
	\includegraphics[width=1\linewidth]{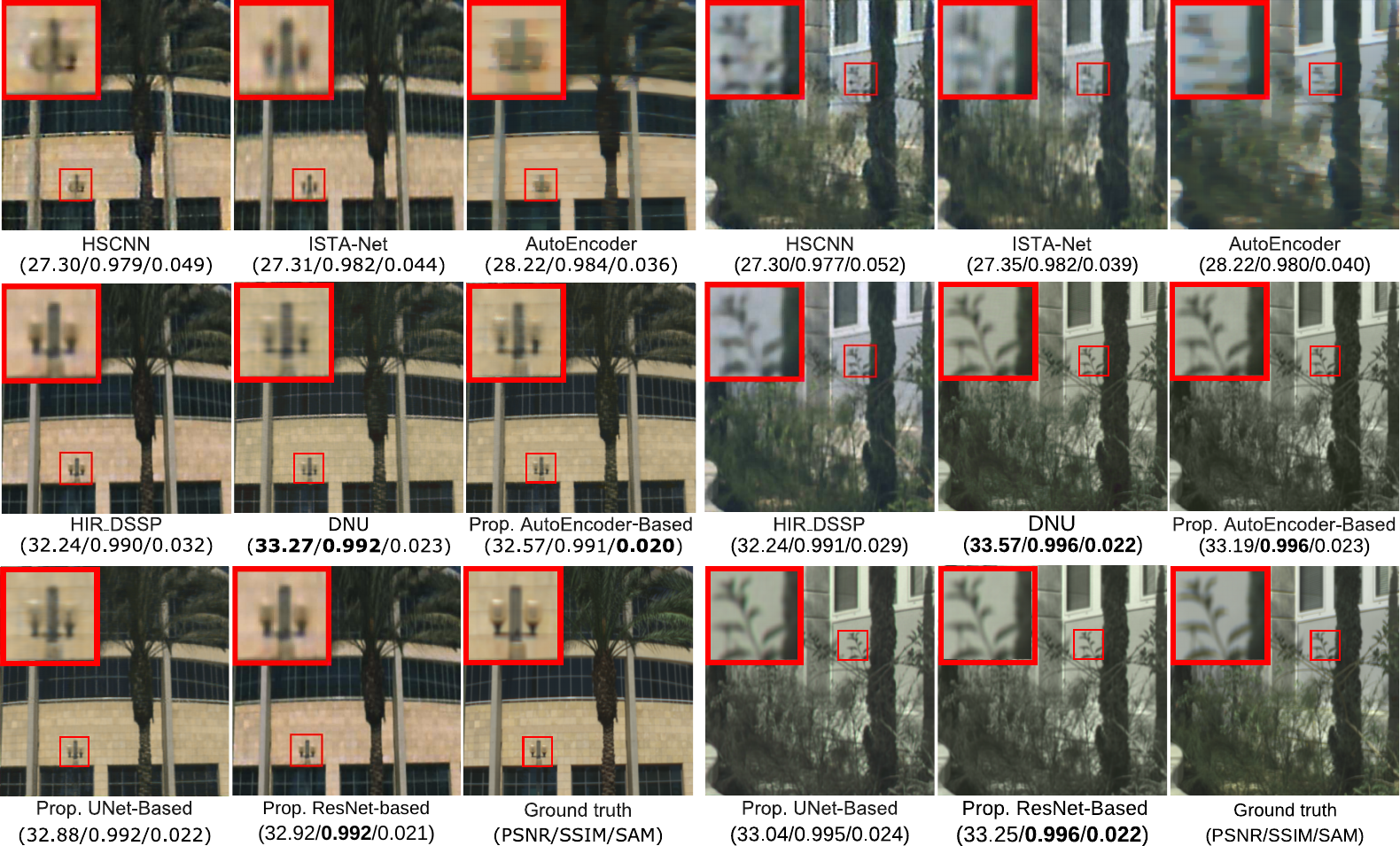}\\ 
	\caption{ Two reconstructed scenes using the 5 learning-based methods  and the \textcolor{black}{three variations of the proposed method, i.e., (AutoEncoder, UNet, and ResNet)-Based.}}
	\label{fig:DataDriven}
\end{figure}
Although the proposed method does not need data to work, this test compares its results with the deep learning approaches to demonstrate the quality achieved. In particular, we use five learning-based methods for comparison: HSCNN \cite{xiong2017hscnn}, ISTA-Net
\cite{zhang2018ista}, Autoencoder \cite{choi2017high}; HIR-DSSP \cite{wang2019hyperspectral} and DNU~\cite{wang2020dnu}. These methods were trained using the public ICVL \cite{arad2016sparse}, Harvard \cite{chakrabarti2011statistics}, and KAIST \cite{choi2017high} hyperspectral image data-sets using their available codes and \textcolor{black}{following the principles in ~\cite{wang2018hyperreconnet,wang2019hyperspectral} to partition the training and testing sets}; the sensing process was evaluated for a single snapshot \textcolor{black}{with 30 dB of SNR}, according to~\cite{wang2020dnu}. \textcolor{black}{For this section, ResNet-based, AutoEnconder-Based, and UNet-based  were used as the Convolutional layer in the proposed method with $\rho=\{0.5, 0.7, 0.7\}$, respectively.} Two testing images of $512\times 512$ of spatial resolution and $31$ spectral bands were chosen to evaluate the different methods, and the reconstruction results and ground truth are shown in Fig. \ref{fig:DataDriven}. It can be observed that the two variants of the proposed method outperform in visual and quantitative results to HSCNN, ISTA-Net, AutoEnconder, HIR-DSSP, up to $(5/0.030/0.020)$ in terms of (PSNR/SSIM/SAM), respectively, and show comparable/close results with respect to the DNU method, which is the best deep learning method. \textcolor{black}{To make a fair run-time comparison of the different methods, all the recovery approaches were running in an Intel (R) Xeon (R) CPU 2.80 GHz. Additionally, since all deep learning methods are implemented to use GPU, we also run it Google Colab source using an NVIDIA Tesla P100 PCIe 16 GB. Table \ref{table:times} shows the running time for reconstructing one spectral image from the compressive measurements. Notice that the proposed methods are iterative; therefore, we employed 2,000 iterations which offers a stable convergence. Although the execution time to obtain a spectral image is longer than most deep learning methods}, the proposed methods have the advantage that it does not require training, i.e., only the compressive measurements are available for the proposed approach.
\begin{table}[!t]
	\setlength\tabcolsep{0.08cm}
	\renewcommand{\arraystretch}{1.0}
	\caption{ Computational complexity of the deep learning and the proposed methods measured as mean time in seconds of 5 trials. }
	\vspace{-0.8em}
	\centering	\resizebox{10.35cm}{!} {
			 \begin{tabular}{|c|c|c|c|c|c|c|c|c|}
        	\hline
			\hline
			\small Methods              & \small HSCNN                     & \small ISTA-Net & \small AutoEncoder                  & \small HIR-DSSP                  & \small DNU                & \multicolumn{1}{c|}{\begin{tabular}[c]{@{}c@{}} \small Prop.\\ \small AutoEncoder\end{tabular}} & \multicolumn{1}{c|}{\begin{tabular}[c]{@{}c@{}} \small Prop.\\ \small  UNet\end{tabular}} & \multicolumn{1}{c|}{\begin{tabular}[c]{@{}c@{}}\small  Prop.\\ \small  ResNet\end{tabular}}  \\ \hline \hline
			GPU Time [s] & 8.708 & \underline{3.224}   &   575.421  &  8.397  &  \textbf{2.744} & 137.375  &  278.0411   & 135.834 \\ \hline \hline
				CPU Time [s] & 72.174 & \underline{27.154}   &   3948.421  &  68.214  &  \textbf{20.727} & 1084.154  &  2224.145  & 997.156  \\ \hline \hline
        \end{tabular}\label{table:times}}
\end{table}

\begin{table}[!t]
	\setlength\tabcolsep{0.08cm}
	\renewcommand{\arraystretch}{1.0}
	\caption{Mean performance comparison for the different recovery methods varying the number of snapshots and noise in SNR dB.}
	\vspace{-0.8em}
	\centering	\resizebox{12.35cm}{!} {
			 \begin{tabular}{|c|c|c|c|c|c|c|c|c|}
        	\hline
			\hline
			\small Shots              & \small Noise                     & \small Metrics & \small GPSR                  & \footnotesize ADMM                  & \footnotesize CSALSA                & \multicolumn{1}{c|}{\begin{tabular}[c]{@{}c@{}}PnP\\ \footnotesize ADMM\end{tabular}} & DIP & Prop. \\ \hline \hline
			\multirow{9}{*}{1} & \multirow{3}{*}{$\infty$} & \small PSNR    &     25.66 $\pm$ 1.780   &  24.32 $\pm$ 1.795  &  25.59 $\pm$ 1.543 & \underline{28.99} $\pm$ 1.642 &  27.93 $\pm$ 2.013   & \textbf{30.92} $\pm$ 1.862 \\ \cline{3-9} 
			&    & \small SSIM    &    0.701  $\pm$ 0.026  & 0.726  $\pm$ 0.019 & 0.790 $\pm$ 0.009 & \underline{0.860} $\pm$ 0.010 &  0.766 $\pm$ 0.023  & \textbf{0.874} $\pm$ 0.018    \\ \cline{3-9} 
			&   &\small 
			SAM     & 0.145  $\pm$ 0.092  &  0.108 $\pm$ 0.101  &     0.152 $\pm$ 0.074   &\underline{0.060} $\pm$ 0.050 &  0.089 $\pm$ 0.074  & \textbf{0.055} $\pm$ 0.061 \\ \cline{2-9} 
			& \multirow{3}{*}{30}       & \small PSNR    &   25.52 $\pm$ 1.802    & 22.68 $\pm$ 1.850  &25.46 $\pm$ 1.842 & \underline{28.82} $\pm$ 1.645 & 27.19 $\pm$ 2.142 & \textbf{29.29} $\pm$ 1.952 \\ \cline{3-9} 
			&                           & \small SSIM    &    0.699 $\pm$ 0.028   &   0.653 $\pm$ 0.021  &  0.701 $\pm$ 0.011  &\underline{0.844} $\pm$ 0.012 &  0.772 $\pm$ 0.029  &    \textbf{0.864}    $\pm$ 0.024  \\ \cline{3-9} 
			&                           & \small SAM     & 0.156 $\pm$ 0.105  &   0.112   $\pm$ 0.108  &    0.167  $\pm$ 0.025  &\underline{ 0.073}  $\pm$ 0.082  & 0.089  $\pm$ 0.092  & \textbf{ 0.062} $\pm$ 0.072 \\ \cline{2-9} 
			& \multirow{3}{*}{20}       & \small PSNR    & 24.67 $\pm$ 1.834 & 21.45 $\pm$ 1.881 & 22.19 $\pm$ 1.872 & 25.42 $\pm$ 1.649	&  \underline{27.53} $\pm$ 2.184 & \textbf{27.94} $\pm$ 1.994 \\ \cline{3-9} 
			&                           &\small SSIM    & 0.682 $\pm$  0.031 & 0.625 $\pm$ 0.028 &  0.672 $\pm$ 0.012 & 0.713 $\pm$ 0.014	 &  \underline{0.783 }  $\pm$ 0.030&    \textbf{0.794} $\pm$ 0.026   \\ \cline{3-9} 
			&                           &\small SAM     & 0.210 $\pm$ 0.111 &  0.220 $\pm$  0.110 &   0.195 $\pm$ 0.031 &0.138  $\pm$ 1.658&  \underline{0.084} $\pm$ 2.214  &  \textbf{ 0.080 } $\pm$ 1.998 \\ \hline 
			\multicolumn{3}{|c|}{CPU Time [s]} & \underline{288.488}$\pm$ 3.142 & 438.812 $\pm$ 3.25 & 308.452 $\pm$ 2.954 & \textbf{198.245} $\pm$ 2.941 & 702.245 $\pm$ 3.154 & 773.235 $\pm$ 3.054 \\ \hline\hline
			
        \end{tabular}\label{table:Robust_analysis}}
\end{table}

\subsection{Robustness Analysis}
\begin{figure}[!t]
	\centering
	\includegraphics[width=1\linewidth]{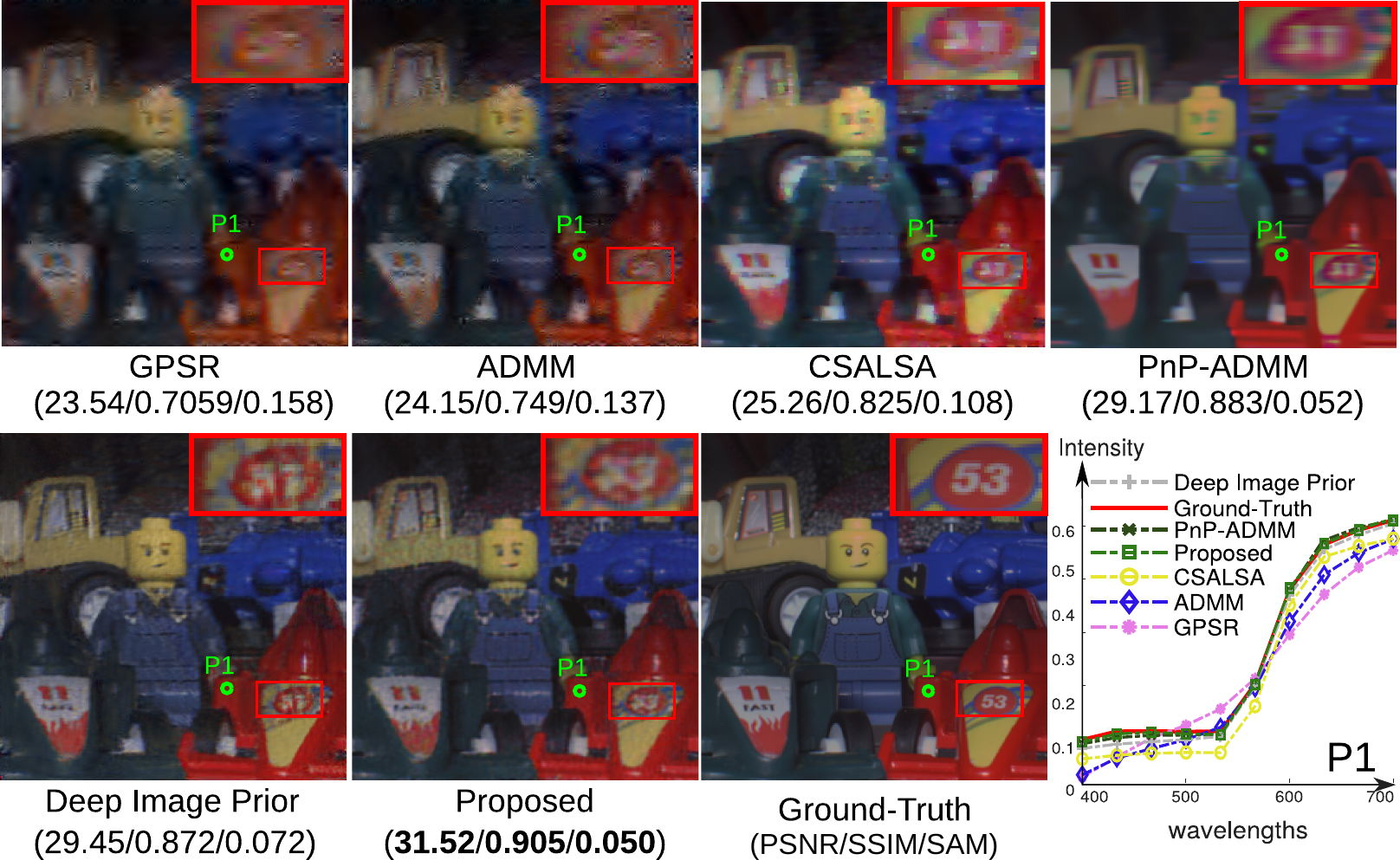}\\\vspace{0.3em}
	\vspace{0.4ex}
	\includegraphics[width=1\linewidth]{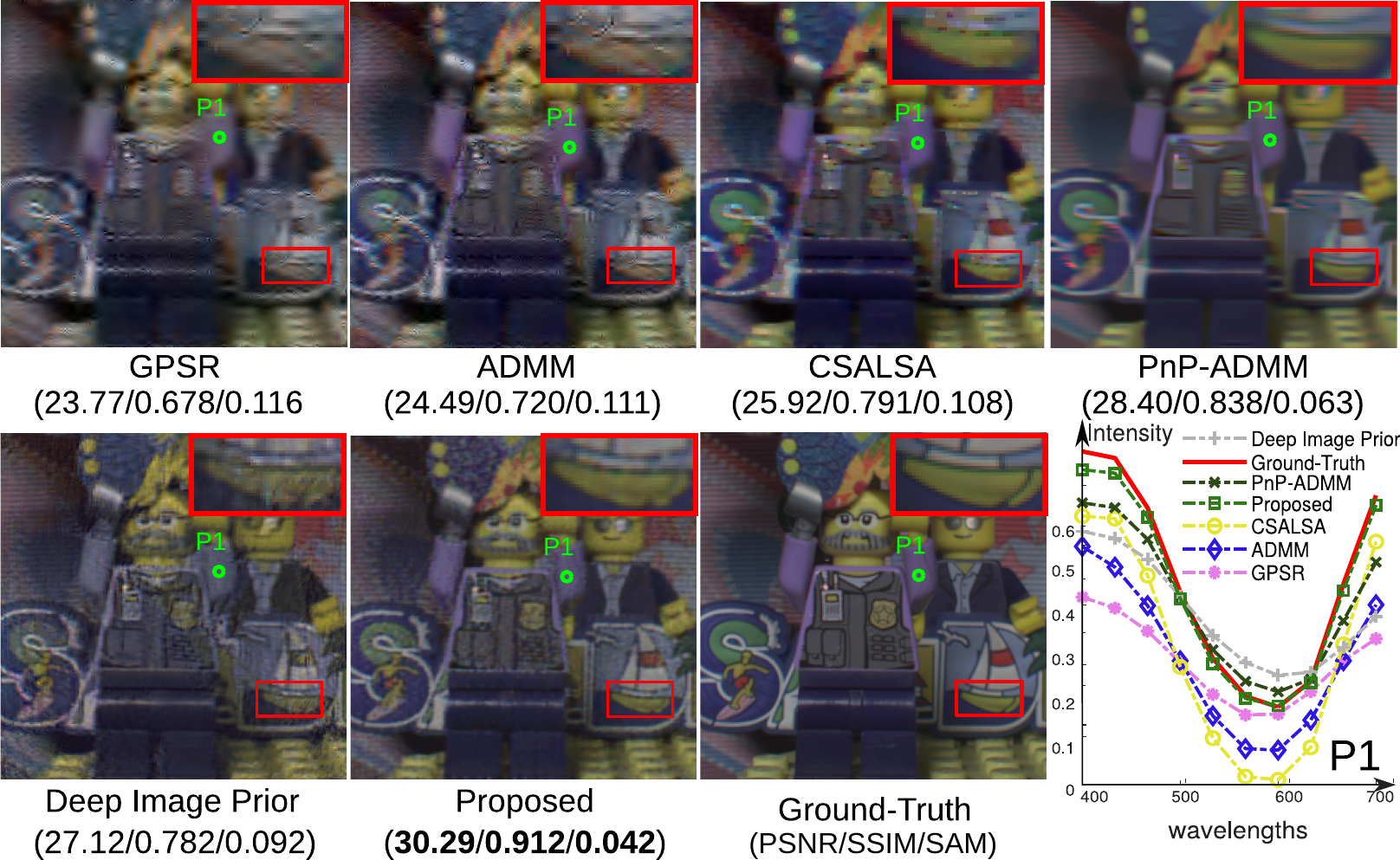}
	\caption{ Two RGB false color reconstructed scenes using the non-data driven methods and the proposed method with its respective metrics are presented. Additionally, the ground-truth and a spectral point of each scene is shown.}
	\label{fig:Nondatadependet}
\end{figure}

Numerical simulations were conducted to demonstrate the robustness of the proposed method at different levels of additive Gaussian noise and the number of snapshots, using the two spectral image obtained in \cite{marquez2020compressive}. \textcolor{black}{Deep learning methods are not flexible to changes in the input, such as the number of spectral bands, and also,} the distribution of training and test data must be similar to obtain good results, for this reason, in this experiment, the proposed method was compared with the state-of-art non-data driven methods. Specifically, we ccompare the proposed method with the GPSR~\cite{figueiredo2007gradient}, using the sparsity assumption in the Wavelet Kronecker Discrete Cosine transform implemented as in~\cite{arguello2014colored}, ADMM~\cite{boyd2011distributed} using the low-rank prior implemented as in~\cite{bacca2019noniterative}, CSALSA~\cite{csalsa} using the 3D total variation, PnP-ADMM~\cite{yuan2020plug} using the BM3D as denoiser, and Deep Image Prior~\cite{ulyanov2018deep} using the ResNet-based network. Three different noise levels were evaluated: 20, 30 dB of signal to noise ratio (SNR) and noiseless case that results in $\infty$ dB. \textcolor{black}{Further, a single CASSI shot was used, which is the extreme case in terms of compression ( See Supplementary Material for a detailed experiment varying the number of snapshots).} \textcolor{black}{Section 5.1 and 5.2 show that the ResNet-based method obtains a slight improvement compared with the proposed UNet-based and AutoEncoder-based. For that reason,} the ResNet-based network was used as the ``Convolutional layers" in the proposed model for this experiment, and the rank factor was fixed as $\rho=0.5$ and $\rho=0.4$ for the DataSet 1 and DataSet 2, respectively. Table \ref{table:Robust_analysis}, presents a comparison of the performance in terms of PSNR, SSIM, and SAM metrics, for the different methods (the results are the average of the two DataSet). Boldface indicates the best result for each case, and the second-best result is underlined. From the Table \ref{table:Robust_analysis}, it can be seen that the proposed method outperforms in almost all cases the other methods. \textcolor{black}{Furthermore, the proposed method shows good noise robustness compared to the other approaches since the proposed method results obtained with 20 SNR improve the other recovery quality, even for the noiseless cases.} 
\textcolor{black}{Notice that the proposed non-training data method obtains good results at the cost of a longer execution time, as reported in Table \ref{table:Robust_analysis}.} 

To visualize the reconstructions and analyze the results in more detail, Figure \ref{fig:Nondatadependet} shows an RGB false color for the reconstruction of each method, for 30 dB of SNR.
Note, that the proposed method, in the zoomed insets, is much cleaner than its counterparts. Additionally, to see the behavior, a single spatial point of each reconstruction for the two Datasets are also presented in Figure \ref{fig:Nondatadependet}. It can be seen that the spectral signatures obtained by the proposed method closely resemble the ground-truth.

\subsection{Validation in a Real  Testbed  Implementation}
This section evaluates the proposed method with real measurements acquired using a testbed implementation.  For this section, the ResNet-based model was used with ($\rho=0.4$), and learning rate $1e-3$. Specifically, two different scenarios of compressed projections were assessed, which are described as follows.

\begin{figure}[b!]
	\centering
	\includegraphics[width=0.8\linewidth]{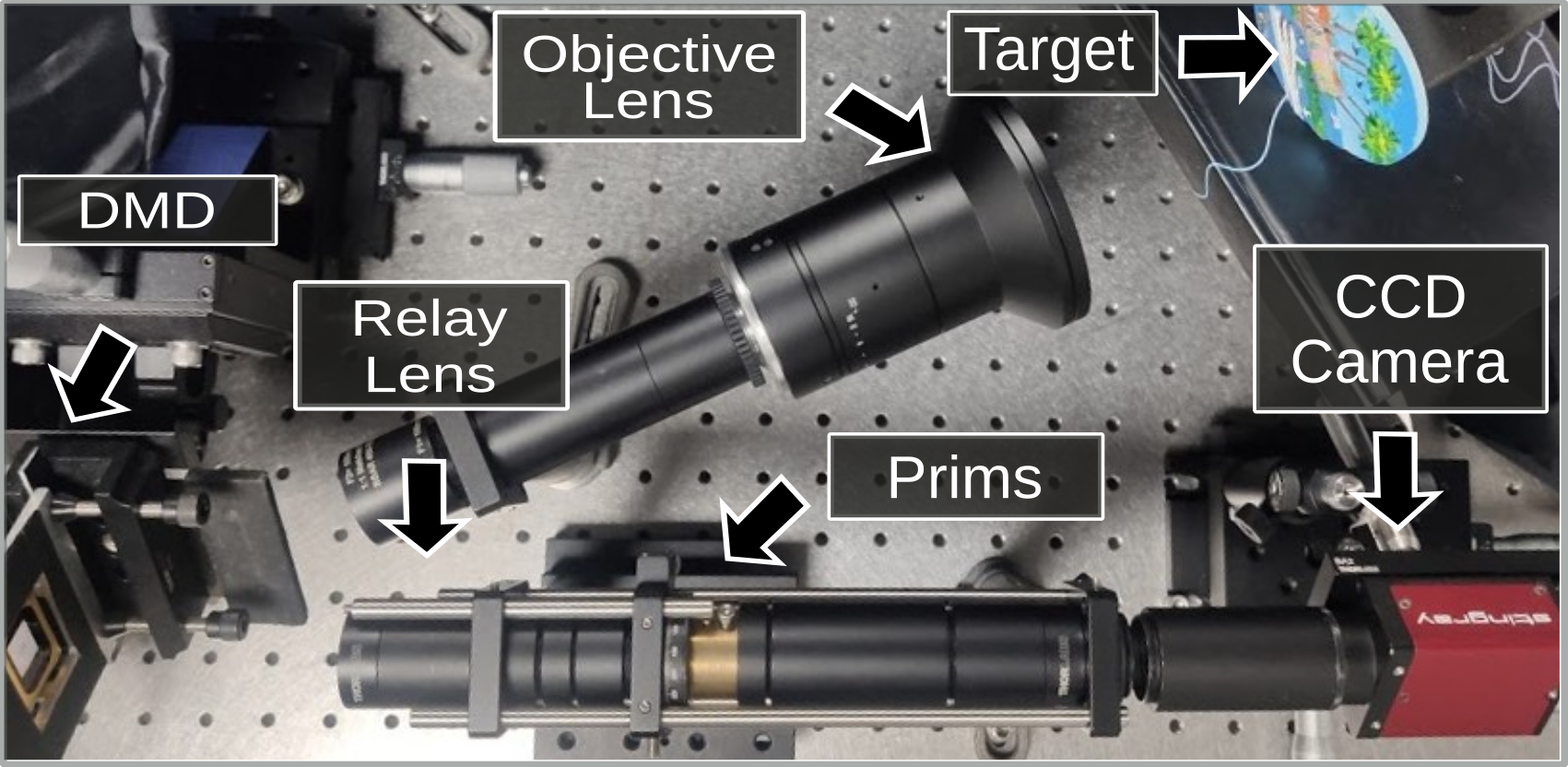}
	\caption{Testbed CASSI implementation where the relay lens focuses the encoded light by the DMD into the sensor after dispersed by the prism.}
	\label{fig:realsetup}
\end{figure}

\subsubsection{Binary Coded Aperture}
\label{sec:binary_real_Data}
This scenario was carried out for one snapshot of the CASSI testbed laboratory implementation depicted in Fig. \ref{fig:realsetup}. This setup contains a $100$-$nm$ objective lens, a high-speed digital micro-mirror device (DMD) (Texas Instruments-DLI4130), with a pixel size of $13,6 \mu m$, where the CA is implemented, an Amici Prism (Shanghai Optics), and a CCD (AVT Stingray F-145B) camera with spatial resolution $1388 \times 1038$, and pitch size of $6.45 \mu m$. The CA spatial distribution for the snapshot comes from blue noise patterns, i.e., this CA is designed according to \cite{correa2016spatiotemporal}. \textcolor{black}{The coding and the scene were implemented to have a spatial resolution of $512\times 512$ pixels and $L=13$ as the resolvable bands}. Notice that the robustness analysis summarized in Table \ref{table:Robust_analysis}, showed that the three best recovery methods were the PnP-ADMM, DIP, and the proposed method; therefore, we decided also to compare them using this real data.

\begin{figure}[!t]
	\centering
	\includegraphics[width=0.9\linewidth]{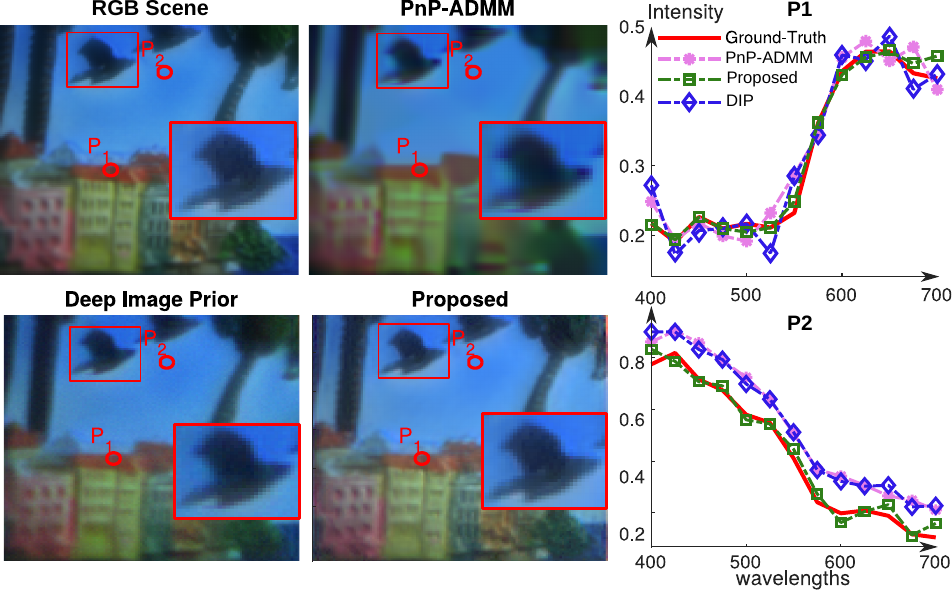}	
	\caption{ (Left) RGB visual representation of the scene obtained with the different methods, (Right),  two spectral signatures of the recovered scenes.}
	\label{fig:CASSI_real_bin}
\end{figure}

\begin{figure}[!b]
	\centering
	\includegraphics[width=0.9\linewidth]{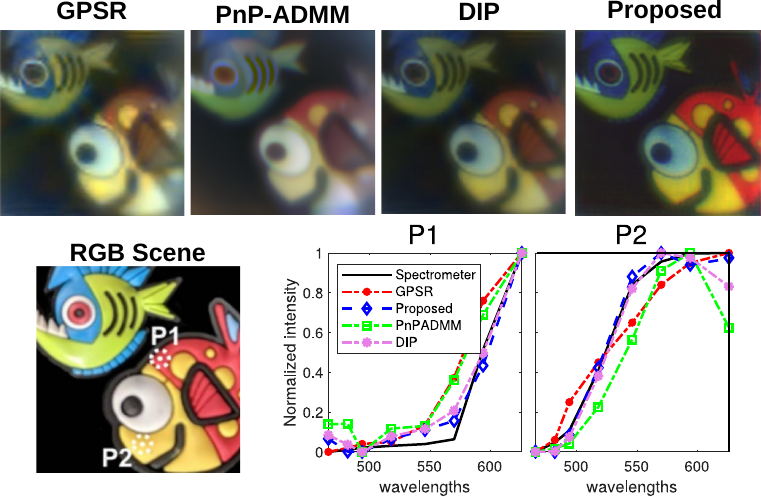}
	\caption{ (Top) RGB visual representation of the scene obtained with the \textcolor{black}{different methods} and the proposed method, (Bottom), \textcolor{black}{RGB scene,} and  normalized spectral signatures of the recovered scenes.}
	\label{fig:CASSI_real_color}
\end{figure}

Figure \ref{fig:CASSI_real_bin} presents the RGB scene obtained with a traditional camera, and the false-colored RGB images corresponding to reconstructed spectral images using the different solvers. Furthermore, the spectral responses of two particular spatial locations in the scene, indicated as red points in the images, are also included and compared with the spectral behavior using a commercially available spectrometer (Ocean Optics USB2000+). The visual results show that the proposed method yield better spatial and spectral reconstruction since the RGB reconstructed is sharper in the proposed scheme, and the spectral signatures are closer to those taken by the spectrometer, this is, the SAM of the normalized signatures obtained from the PnP-ADMM algorithm is 0.188, Deep Image Prior is 0.205, and the SAM associated to the proposed method is 0.120. These numerical results validate the performance of the proposed method with real data for a real CASSI setup using a binary-coded aperture.
\subsubsection{Colored Coded Aperture}

The real data for this second test was provided by~\cite{galvis2019shifting}. In particular, the main difference with the data of Section \ref{sec:binary_real_Data} is that the spatial modulation is a Colored CA, where each pixel can be seen as a filter with its spectral response, (further details regarding Colored CA can be found in \cite{arguello2014colored,galvis2019shifting}). The optical elements in this testbed implementation were the same used in the previous setup, where the DMD was used to emulate the Colored CA. The coding and the scene were implemented to have a spatial resolution of $256\times 256$ pixels and $L=8$ as the resolvable bands, \textcolor{black}{where two shots were chosen. The work in} \cite{galvis2019shifting} uses a hand-crafted method, which does not require training data, and the GPSR algorithm was used as a recovery algorithm; therefore, the proposed method was compared with this method, \textcolor{black}{the DIP, and the PnP-ADMM methods. } Figure \ref{fig:CASSI_real_color} (Top) shows the RGB mapping of the recovered scenes. There, it can be seen that the proposed method provides a cleaner version of the scene. Additionally, two spatial points were chosen to evaluate the spectral behavior illustrated in Figure  \ref{fig:CASSI_real_color} (Bottom). It can be seen that the spectral signature provided by the proposed method is closer to the obtained with the spectrometer compared with the other methods, in fact, the SAM of the normalized signatures obtained from the GPSR algorithm is 0.120 and the SAM associated to the proposed method is 0.057. These results validate the effectiveness of the proposed method on real data for two variations of CASSI systems.

\section{Conclusions}
A method for reconstructing spectral images from the CSI measurements has been proposed. The proposed scheme is based on the fact that the spectral images can be generated from a convolutional network whose input features comes from a low-rank Tucker representation. Although the proposed method is based on a convolutional network framework, it does not require training data, only the compressed measurements. This method was evaluated in three scenarios: noiseless, noisy, and real data implementation. In all of them, the proposed method outperforms the image quality reconstruction compared with state-of-the-art methods. In particular, the proposed method with 20 SNR levels of noise in the CSI measurements outperforms its counterparts in up to 4 dB in the PSNR measure. \textcolor{black}{Although the proposed method was tested in two real CSI measurements, these toy scenes contain piece-wise constant regions, which are not common in real-life scenes. Therefore, we will consider evaluating more realistic CSI measurements as used in Section 5.1 in future works. Furthermore, the proposed methods can be extended and used in others compressive systems where the data set is limited.}

\section{Acknowledgments}
Universidad Industrial de Santander under VIE-project 2699.

\section{Disclosures}
 The author declares no conflicts of interest.


\bibliography{sample}






\end{document}